%% file: main.tex
\documentclass[11pt]{article}

\usepackage[preprint]{acl}
\usepackage{times}
\usepackage{latexsym}
\usepackage[T1]{fontenc}
\usepackage[utf8]{inputenc}
\usepackage{microtype}
\usepackage{inconsolata}
\usepackage{booktabs}
\usepackage{amsmath}
\usepackage{amssymb}
\usepackage{tabularx}
\usepackage{graphicx}
\usepackage{subcaption}
\usepackage{adjustbox}
\usepackage{tcolorbox}
\usepackage{listings}
\title{FinReflectKG - EvalBench: Benchmarking Financial KG with Multi-Dimensional Evaluation}

\author{
  Fabrizio Dimino \\
  Domyn \\
  New York, US \\
  \And
  Abhinav Arun \\
  Domyn \\
  New York, US \\
  \And
  Bhaskarjit Sarmah \\
  Domyn \\
  Gurugram, India \\
  \And
  Stefano Pasquali \\
  Domyn \\
  New York, US \\
}

\begin{document}

\maketitle
\input{sections/abstract}

\input{sections/introduction}
\input{sections/related_work}
\input{sections/methodology}

\input{sections/results}
\input{sections/conclusion}

% References
%%\bibliographystyle{acl_natbib}
\bibliography{references}

\clearpage
\appendix
\input{sections/appendix}

\end{document}

%% file: sections/abstract.tex
\begin{abstract}
Large language models (LLMs) are increasingly used to extract structured knowledge from unstructured financial text, yet the field lacks a standardized benchmark and unified evaluation protocol for financial knowledge graphs (KGs). We introduce \textbf{FinReflectKG-EvalBench}, a benchmark and bias-aware evaluation framework for KG extraction from SEC 10-K filings. Building on FinReflectKG \cite{arun2025finreflectkgagenticconstructionevaluation}, EvalBench supports single-pass, multi-pass, and reflection-based extraction, and evaluates candidate triples with a commit-then-justify LLM-as-Judge ensemble equipped with explicit bias controls (position effects, leniency, verbosity, and world-knowledge reliance). Judges assign binary labels for faithfulness, precision, and relevance, while comprehensiveness is assessed at the chunk level using a three-level scale. Inter-judge agreement is high across dimensions, supporting that the protocol yields stable and auditable judgments.
Empirically, reflection achieves the strongest performance on comprehensiveness, precision, and relevance, while single-pass maintains the highest faithfulness.
\\[2mm]
\textbf{Keywords:} Knowledge Graphs, LLM-as-Judge, Evaluation, SEC Filings, Large Language Models, Financial AI
\end{abstract}

%% file: sections/introduction.tex
\section{Introduction}
Large language models (LLMs) are increasingly used in finance to process long, complex, and unstructured documents such as regulatory filings, earnings reports, and legal disclosures. A common target representation for such extracted information is the Knowledge Graph (KG), which structures textual content into relational triples and enables applications including compliance monitoring, risk assessment, investment research, and large-scale financial analytics~\cite{li2024findkg, li2024graphreader}.

Recent advances in LLM prompting strategies, such as multi-turn extraction, self-reflection, and iterative refinement, have substantially improved the quality of automatically extracted triples~\cite{zhang2024extract}. However, progress in extraction capability has outpaced progress in evaluation. In high-stakes financial settings, even subtle extraction errors can propagate downstream, leading to incorrect risk signals, misleading investment insights, or regulatory exposure. As a result, the lack of reliable, reproducible, and bias-aware evaluation protocols for financial KG extraction remains a critical bottleneck.

Evaluating KG extraction is inherently challenging. Financial documents are dense, context-dependent, and often require domain expertise to interpret correctly. Gold-standard annotations are expensive and difficult to scale, while exact matching metrics fail to capture semantic correctness, partial entailment, or relevance. Consequently, recent work has increasingly relied on LLM-as-Judge frameworks as a scalable alternative to human evaluation~\cite{li2024llmsasjudgescomprehensivesurveyllmbased, evidently2025guide}. Despite their practical appeal, LLM judges introduce a new set of risks: they are sensitive to prompt phrasing and answer ordering~\cite{shi2024judgingthejudges}, often overly lenient~\cite{ye2024justice}, and prone to stylistic or persuasive biases that reward fluency over factual grounding~\cite{hwang2025trickgrader}.

These limitations are especially problematic in the context of triple-level KG extraction, where evaluation requires fine-grained judgments about faithfulness to source text, factual precision, and relevance to a predefined schema. While several open-source evaluation toolkits provide LLM-based evaluators for tasks such as question answering and summarization~\cite{deepeval,ragas,trulens,langsmith}, they are not designed to address the unique challenges posed by structured KG extraction in regulated domains such as finance.

To address this gap, we introduce \textbf{FinReflectKG-EvalBench}, a benchmark for evaluating financial KG extraction from SEC 10-K filings. Building on the agentic construction principles of FinReflectKG~\cite{arun2025finreflectkgagenticconstructionevaluation}, our benchmark integrates schema-aware extraction with a conservative, bias-aware evaluation protocol. Unlike prior work, FinReflectKG-EvalBench systematically compares multiple extraction strategies—single-pass, multi-pass, and reflection-based extraction—under a unified evaluation framework tailored to financial risk-sensitive applications.

This paper makes the following contributions:
\begin{enumerate}
    \item We introduce \textbf{FinReflectKG-EvalBench}, the first benchmark for financial KG extraction from SEC filings with a reproducible and bias-aware evaluation protocol.
    \item We propose a heterogeneous ensemble of LLM judges that aggregates decisions from multiple models to mitigate prompt sensitivity, positional bias, and evaluation leniency.
    \item We present a systematic comparison of extraction strategies across four complementary dimensions: faithfulness, precision, relevance, and comprehensiveness.
    \item We quantify evaluation stability via inter-judge agreement (Krippendorff's $\alpha$), supporting that the protocol yields consistent judgments across dimensions.
\end{enumerate}

%% file: sections/related_work.tex
\section{Related Work}
\subsection{Knowledge Graph Construction from Financial Documents}

Financial knowledge graphs have been constructed from a wide range of real-world artifacts, spanning regulatory filings, annual reports, brokerage research, and financial news. A recurring theme across these efforts is that financial text is both highly structured (tables, sections, standardized disclosures) and semantically ambiguous (domain jargon, entity aliases, evolving corporate structures), which makes schema design and normalization central to KG quality~\cite{li2024findkg, li2024graphreader}. In practice, many pipelines therefore adopt schema, or ontology, driven representations to constrain extraction and improve interoperability, building on standardized vocabularies such as the Financial Industry Business Ontology (FIBO) and related financial ontologies~\cite{edmcouncil_fibo, bennett2013fibo}.

Beyond filings, several works mine analyst and brokerage research reports to build domain-specific KGs that capture entities, events, and relations useful for downstream analytics. For instance, FR2KG provides a dataset and an evaluation setup for automated construction of a financial research report knowledge graph~\cite{wang2021fr2kg}; while other work demonstrates scalable KG construction by mining massive brokerage research corpora, emphasizing template design, relation canonicalization, and quality control under limited supervision~\cite{cheng2022democratizing}. Complementary lines of research explore knowledge discovery and graph extraction from financial texts with minimal or no labeled data, highlighting the feasibility of end-to-end pipelines but also exposing sensitivity to document heterogeneity and extraction noise~\cite{oksanen2022graph}.

More recently, LLM-driven information extraction has expanded the design space from strictly closed schemas to more open or semi-open extraction regimes; GoLLIE shows how guideline-following instruction tuning can improve zero-shot information extraction under complex annotation schemas~\cite{sainz2024gollie}. This is particularly relevant for financial filings, where open IE over long documents can surface diverse relations that are hard to enumerate a priori; for example, recent work constructs KGs from 10-K reports using open extraction guided by relation families (e.g., business, transaction, personnel) to support downstream financial applications~\cite{guo2024open10k}. In parallel, temporal and dynamic KGs built from financial news model how entities and relations evolve over time, enabling time-aware downstream tasks such as anomaly detection and link prediction~\cite{li2024findkg}.

As extraction pipelines become more capable, they also become more interactive and iterative. Recent prompting strategies aim to improve triple quality by allowing models to revise earlier outputs in light of schema constraints or source-text inconsistencies~\cite{papaluca2024zero, mou2024llmkg}. Agentic frameworks further extend this paradigm by decomposing extraction into coordinated sub-tasks, such as entity identification, relation grounding, and consistency checking~\cite{arun2025finreflectkgagenticconstructionevaluation}.

\subsection{Evaluation of Knowledge Graph Extraction}

Evaluating the quality of knowledge graph (KG) extraction remains an open and multifaceted challenge. Traditional metrics such as exact triple matching or surface-level overlap are brittle: they fail to account for semantic equivalence, partial correctness, or contextual relevance, and often penalize correct extractions that differ in surface form from a gold standard~\cite{chang2024survey}. Moreover, metrics originally developed for ontology alignment or schema mapping emphasize lexical similarity and are insufficient for capturing deeper semantic relationships between extracted facts~\cite{seo2022structural}.

Beyond matching metrics, research in KG quality assessment has proposed structural quality metrics-such as connectivity, consistency, coverage-that evaluate how well a constructed KG satisfies desirable properties such as coherence and completeness across the graph, independent of surface forms~\cite{wang2021knowledge}. 

Human annotation remains the gold standard for extraction evaluation due to its ability to interpret nuanced semantics and domain-specific knowledge. However, manual annotation is costly, slow, and difficult to scale, particularly for specialized domains such as finance that require expert judgment. To address this, numerous works have employed semantic similarity and embedding-based comparisons between extracted triples and references~\cite{wei2025semantic_kg}. These approaches use vector representations to measure semantic closeness and have been applied both for entity alignment and relation matching, but they struggle with interpretability at the schema level and may conflate distinct semantic errors.

As a result, evaluation pipelines across both information extraction and KG extraction are increasingly turning to LLM-based judges as a pragmatic compromise between scalability and semantic sensitivity. LLM-as-Judge frameworks leverage generative models to interpret and compare extracted structures in natural language, allowing for richer criteria such as faithfulness, relevance, and factuality to be incorporated into automated evaluation~\cite{li2024from}. LLM judges can also produce structured critiques or rationale that assist error analysis, despite introducing their own challenges in prompt design, consistency, and calibration.

\subsection{LLM-as-Judge: Reliability and Biases}
A growing literature studies LLM-based evaluators not only as a convenient proxy for human assessment, but as a complex measurement instrument whose outputs depend on the evaluation protocol, prompt specification, and judge model choice. In practice, many widely-used evaluation pipelines adopt pairwise comparisons to reduce score-scale drift and improve reproducibility~\cite{zheng2023mtbench}. Closely related efforts such as AlpacaEval operationalize this paradigm for instruction-following evaluation at scale, while explicitly documenting persistent confounders that can distort model rankings~\cite{li2023alpacaeval}.

Beyond protocol choices, recent work shows that LLM-judge reliability is constrained by systematic biases that are often orthogonal to task correctness. Positional effects remain a recurring failure mode: LLM judges can favor candidates based on presentation order rather than content quality, a phenomenon documented both in benchmark-centric analyses~\cite{zheng2023mtbench} and in dedicated studies that quantify position bias across judges, tasks, and comparison settings~\cite{dimino2025bias,shi2024positionbias}. Separate lines of work characterize style-related biases, where superficial attributes-such as fluency, formatting-can outweigh factual accuracy, motivating multi-dimensional rating schemes instead of collapsing evaluation into a single scalar score~\cite{wu2023style}. Complementing these findings, bias taxonomies for judges also include more nuanced effects such as authority, beauty, fallacy-oversight biases under controlled perturbations, showing that even strong judges can be systematically steered~\cite{chen2024judgementbias}.

Importantly, judge behavior can interact with model communication strategies. For example, LLM-judges can penalize epistemic markers and expressions of uncertainty, indicating that evaluation may inadvertently disincentivize calibrated or honest uncertainty reporting~\cite{lee2025uncertainty}. Another recurring concern is self-preference or self-bias, where judges favor outputs that resemble their own generations or systematically assign higher scores to their own responses~\cite{chen2025selfpreference}. These effects suggest that judge selection and ensemble design are not merely engineering details, but central components of evaluation validity.

%% file: sections/methodology.tex
\section{Methodology}

\subsection{Corpus and span construction}
We construct \textbf{FinReflectKG - EvalBench} on the corpus of U.S.\ SEC Form 10-K filings from S\&P~100 companies for fiscal year 2024. Formally, let $\mathcal{D}$ denote this corpus where each document $d\in\mathcal{D}$ is segmented into text spans $\mathcal{X}_d=\{x_{d,1},\dots,x_{d,K_d}\}$ using a deterministic, structure-aware chunking scheme tailored to financial filings. An information extractor $E$ maps each span to a set of candidate triples,
\[
T_x = E(x) \subseteq \mathcal{T}, \qquad 
t = (s, r, o) \in \mathcal{T},
\]
where $s$ and $o$ are subject and object entity mentions, and $r \in \mathcal{R}$ is a relation from a predefined financial vocabulary.

We consider three extraction modes. Single-pass uses a single LLM for both extraction and normalization. Multi-pass splits the task between two LLMs: one extracts triples, while the other normalizes them according to rubric parameters. Reflection employs an agentic iterative workflow, where extraction and feedback loops refine triples until inconsistencies are resolved or a maximum iteration limit is reached.

\subsection{Evaluation dimensions}
Across the three extraction modes, the task is to evaluate the quality of the candidate triples produced for each span $x$ with candidate set $T_x$. Evaluation is conducted along four complementary dimensions. First, \textbf{faithfulness} $F$ measures whether the content of a triple is factually grounded in the source text, without relying on world knowledge or bridging inferences. Second, \textbf{precision} $P$ assesses the clarity and specificity of triples, penalizing generic placeholders (e.g., “Company”) and imprecise expressions of quantities or dates. Third, \textbf{relevance} $R$ checks whether the triple contributes directly to the main theme of the source span rather than introducing tangential information. Finally, \textbf{comprehensiveness} $C$ is measured on a three-level scale (good, partial, poor) and it evaluates coverage at the chunk level, scoring how well the set of triples represents all atomic core facts. 

To aggregate results across the corpus, let $\mathcal{X}=\bigcup_{d\in\mathcal{D}}\mathcal{X}_d$ and $\mathcal{T}_{\text{all}}=\bigcup_{x\in\mathcal{X}}T_x$.  
Local binary metrics (faithfulness, precision, relevance) are micro-averaged:
\[
\bar F=\frac{1}{|\mathcal{T}_{\text{all}}|}\sum_{(x,t)}F(x,t),\qquad
\bar P,\;\bar R \text{ analogously}.
\]  
Comprehensiveness is macro-averaged across spans:
\[
\bar C=\frac{1}{|\mathcal{X}|}\sum_{x\in\mathcal{X}} C(x).
\]

\subsection{LLM-as-Judge evaluation protocol}
We evaluate extracted triples with a heterogeneous LLM-as-Judge ensemble $J$, instantiated with \textsc{Qwen3-235B-A22B}, \textsc{gpt-oss-120B}, and \textsc{Llama-3.3-Nemotron-Super-49B-v1.5}, and configured for deterministic decoding (temperature $=0.0$). The judges are intentionally diverse in architecture and training data, which makes agreement a practical proxy for evaluation stability: when all three judges converge on the same label, we treat the verdict as high-confidence; when they disagree, we flag the instance as ambiguous and route it to human review (human-in-the-loop).

As shown in \autoref{fig:prompt}, and following the commit-then-justify paradigm~\cite{lopez2023can}, the judge first produces a structured verdict and then a concise justification (up to 15 words). To further support error analysis, we introduce a warning signal that highlights extraction errors and provides actionable correction paths. These signals can also be leveraged within a feedback loop for iterative self-improvement.

To ensure reliable and reproducible judgments, we enforce strict bias controls. First, we adopt a principle of conservatism: whenever the evidence is ambiguous, the judge defaults to a negative decision (0), thereby mitigating leniency bias. Second, we enforce locality, strictly prohibiting the use of world knowledge or inferences beyond the provided text. Third, we guarantee position independence by instructing the judge not to let the order or placement of sentences influence its verdicts. Finally, we ensure verbosity independence, so that the length or surface form of a candidate triple does not bias the evaluation outcome. In addition, we improve consistency and calibration by including few-shot examples for each evaluation criterion in the judge’s prompt. Concrete examples are provided in the Appendix.

\begin{figure}[ht]
\centering
\begin{minipage}{0.95\linewidth}
\lstset{language={} }
\begin{lstlisting}
role: "Knowledge Graph Evaluator"
task: "Determine if the triplet is [evaluation criterion] 
       with respect to the source text context"
instructions: 
  Decision rule:
    - Return 1 if [criterion satisfied], 
    - Return 0 if [criterion not satisfied].
  Bias controls:
    - Be conservative: when uncertain, return 0 (leniency bias).
    - Do NOT infer or add information beyond the text (world knowledge bias).
    - Do NOT let sentence position in the source text affect the decision (position bias).
    - Do NOT let the length of the triplets affect the decision (verbosity bias).
  Reasoning vs Warning:
    - Reasoning: Brief explanation of the verdict (up to 15 words).
    - Warning: actionable tag(s) for error type; Do NOT duplicate reasoning; use empty string if no actionable issue.
  Output policy:
    - Valid JSON array only, single line.
    - Each item: {"verdict":0|1,"reasoning":"...",
    "warning":"..."}

    Examples:
    ...
\end{lstlisting}
\end{minipage}
\caption{System prompt design}
\label{fig:prompt}
\end{figure}

\subsection{Inter-judge agreement}
Because our evaluation relies on a heterogeneous LLM-as-Judge panel, disagreement is not treated as noise but as a diagnostic signal: it typically concentrates on borderline cases where evidence is underspecified in the span or where multiple valid interpretations exist under strict locality constraints. To quantify stability across the three judges, we report \textbf{Krippendorff's $\alpha$} for each evaluation dimension, computed over the full set of judged instances. Unlike pairwise Cohen's $\kappa$, $\alpha$ naturally supports multiple annotators and corrects for chance agreement while remaining interpretable under class imbalance.

Formally, for a given metric $m \in \{F,P,R,C\}$ with label space $\mathcal{Y}_m$, Krippendorff's $\alpha_m$ is defined as
\[
\alpha_m = 1 - \frac{D_o}{D_e},
\]
where $D_o$ is the observed disagreement and $D_e$ is the disagreement expected by chance under the empirical label marginals.
For comprehensiveness, we treat the tree outputs as nominal categories, since they represent discrete coverage regimes rather than a calibrated numeric scale.

%% file: sections/results.tex
\section{Results}

Evaluation of triple generation often treats faithfulness, relevance, precision, and comprehensiveness as separate dimensions. While such a decomposition is informative, we argue that these criteria must ultimately be interpreted jointly in order to capture the full spectrum of trade-offs involved in knowledge graph construction. 

\subsection{Performance across extraction modes}
\begin{table}[ht]
\centering
\caption{Evaluation results across extraction modes.}
\label{tab:results}
\begin{tabularx}{\columnwidth}{l *{3}{>{\centering\arraybackslash}X}}
\toprule
 & \textbf{Single} & \textbf{Multi} & \textbf{Refle-} \\
 & \textbf{Pass}   & \textbf{Pass}  & \textbf{ction}\\
\midrule
Comprehensiveness & 58.75 & 66.32 & \textbf{74.27} \\
Faithfulness      & \textbf{85.11} & 75.92 & 84.80 \\
Precision         & 60.61 & 62.31 & \textbf{64.23} \\
Relevance         & 84.35 & 81.39 & \textbf{90.44} \\
\bottomrule
\end{tabularx}
\end{table}

As reported in \autoref{tab:results}, performance varies markedly across extraction modes. The reflection mode achieves superior results in comprehensiveness, precision, and relevance, whereas the single-pass mode yields the highest score in faithfulness. This pattern is consistent with the intuition that reflection, by design, generates a larger set of triples per chunk, thereby capturing a broader range of atomic core facts. From the perspective of knowledge graph construction, this ability to recover a wider coverage of facts is highly desirable, as it directly impacts the downstream utility of the graph. 

At the same time, the relative decline in faithfulness observed for reflection suggests an inherent trade-off: expanding coverage increases the risk of generating triples that extend beyond the strict boundaries of the source text. In contrast, the single-pass approach, though less comprehensive, remains more conservative and better aligned with the original text. 

Turning to precision, we observe that absolute scores remain relatively modest across all modes, indicating the need for further improvements. Nonetheless, reflection achieves the best performance in this dimension, suggesting that iterative reasoning can modestly improve structural accuracy. By contrast, relevance yields consistently higher values across modes, with reflection leading, which indicates that most generated triples-despite occasional issues of faithfulness or precision-remain topically aligned with the source text.

\subsection{Inter-judge agreement}

To assess the stability of these trends under our heterogeneous LLM-as-Judge panel, we report inter-judge agreement using Krippendorff's $\alpha$ (nominal) for each evaluation dimension, computed over the full dataset (no missing labels).

\begin{table}[ht]
\centering
\caption{Inter-judge agreement (Krippendorff's $\alpha$) across evaluation dimensions.}
\label{tab:agreement}
\begin{tabular}{lc}
\toprule
\textbf{Dimension} & $\boldsymbol{\alpha}$ \\
\midrule
Faithfulness      & 0.88 \\
Precision         & 0.84 \\
Relevance         & 0.77 \\
Comprehensiveness & 0.71 \\
\bottomrule
\end{tabular}
\end{table}

Agreement is highest for faithfulness ($\alpha=0.88$), consistent with the criterion largely reducing to verification of explicit textual support. Precision also exhibits strong agreement ($\alpha=0.84$), reflecting that specificity issues are comparatively easier to identify consistently. Agreement is lower for relevance ($\alpha=0.77$) and comprehensiveness ($\alpha=0.71$), which require judgments about salience and coverage and are inherently more sensitive to extraction granularity under strict locality constraints.

Overall, these agreement levels indicate that the evaluation protocol is sufficiently reliable to support comparison across extraction modes in \autoref{tab:results}. Taken together, the results highlight complementary strengths across strategies: reflection provides the strongest coverage and topical alignment and yields modest gains in structural accuracy, while single-pass remains the most conservative in factual grounding.

%% file: sections/conclusion.tex
\section{Conclusions}
We introduced \textbf{FinReflectKG-EvalBench}, a benchmark for financial knowledge graph extraction from SEC 10-K filings paired with a reproducible, bias-aware evaluation framework. Our approach combines schema-aware extraction with a conservative LLM-as-Judge protocol, explicit bias controls, and a deterministic commit-then-justify procedure with actionable warning signals.

Empirically, we find a clear trade-off across extraction strategies: reflection performs best on comprehensiveness, precision, and relevance, while single-pass achieves the highest faithfulness. Inter-judge agreement is substantial across dimensions, supporting that the proposed evaluation protocol yields stable and auditable judgments for comparing extraction methods.

Overall, FinReflectKG-EvalBench enables fine-grained, multi-dimensional benchmarking of financial KG extraction and provides practical evaluation primitives that facilitate transparent error analysis and targeted human review of ambiguous cases.

\section*{Limitations}
FinReflectKG-EvalBench is currently scoped to U.S.\ SEC Form 10-K filings from S\&P~100 companies for fiscal year 2024, which limits generalization to other issuers, time periods, jurisdictions, languages, and document types (e.g., 10-Q, 8-K, earnings call transcripts, and contracts). Extending the benchmark across corpora and time is necessary to assess robustness under domain shift and temporal drift.

Our evaluation relies on an LLM-as-Judge ensemble. Although we enforce deterministic decoding, explicit bias controls, and observe substantial inter-judge agreement, LLM judges may still exhibit residual failure modes. Agreement should therefore be interpreted as protocol stability rather than ground-truth correctness; targeted human audits remain important for high-stakes deployments.

Finally, our metrics capture complementary aspects of extraction quality but do not fully characterize downstream utility. In particular, chunk-level comprehensiveness does not measure document-level coverage, cross-chunk consistency, or entity resolution over entire filings. Future work should incorporate graph consistency checks, entity linking/normalization evaluation, and task-based metrics tied to downstream financial applications.

%% file: sections/appendix.tex
\section{Appendix}

\label{app:faithfulness}
\begin{tcolorbox}[title=Faithfulness Examples]
\emph{Source Text:} "OpenAI signed a \$1B deal with Microsoft in 2024 in Texas" \\

\textbf{Supported (1):} \\
\emph{Triplet:} ["OpenAI","Signed","\$1B deal"] \\
\emph{Verdict:} 1 (Supported), \emph{Reasoning:} Triplet grounded in source text, \emph{Warning:} None \\

\textbf{Not Supported (0):} \\
\emph{Triplet:} ["OpenAI","Buy","Microsoft"] \\
\emph{Verdict:} 0 (Not Supported), \emph{Reasoning:} Triplet not grounded in source text, \emph{Warning:} Possible hallucination
\end{tcolorbox}

\label{app:precision}
\begin{tcolorbox}[title=Precision Examples]
\emph{Source Text:} "OpenAI signed a \$1Billion deal with Microsoft" \\

\textbf{Precise (1):} \\
\emph{Triplet:} ["OpenAI","Partners\_With","Microsoft"] \\
\emph{Verdict:} 1 (Precise), \emph{Reasoning:} Specific entities and relation, \emph{Warning:} None \\

\emph{Triplet:} ["OpenAI","Signed","\$1Billion deal"] \\
\emph{Verdict:} 1 (Precise), \emph{Reasoning:} Specific entities and amount, \emph{Warning:} None \\

\textbf{Not Precise (0):} \\
\emph{Triplet:} ["Company","Related\_To","Something"] \\
\emph{Verdict:} 0 (Not Precise), \emph{Reasoning:} Generic entity and broad relation, \emph{Warning:} Generic entity \\

\emph{Triplet:} ["OpenAI","Signed","\$2Billion deal"] \\
\emph{Verdict:} 0 (Not Precise), \emph{Reasoning:} Amount mismatch with text, \emph{Warning:} Amount mismatch
\end{tcolorbox}

\label{app:relevance}
\begin{tcolorbox}[title=Relevance Examples]
\emph{Source Text:} "OpenAI signed a \$1Billion deal with Microsoft in Texas" \\

\textbf{Relevant (1):} \\
\emph{Triplet:}["OpenAI","Partners\_With","Microsoft"] \\
\emph{Verdict:} 1 (Relevant), \emph{Reasoning:} Triplet relevant for the source text, \emph{Warning:} None

\textbf{Not Relevant (0):} \\
\emph{Triplet:} ["OpenAI","Signed","in Texas"] \\
\emph{Verdict:} 0 (Not Relevant), \emph{Reasoning:} Location not relevant to the main topic, \emph{Warning:} Off-topic
\end{tcolorbox}

\label{app:comprehensiveness}
\begin{tcolorbox}[title=Comprehensiveness Examples]
\emph{Source Text:} "In 2024, OpenAI signed a \$1Billion deal with Microsoft for AI partnership in Texas." \\

\textbf{Score 3 (Good):} \\
\emph{Set of Triplets:} 
["OpenAI","Partners\_With","Microsoft"], 
["OpenAI","Signed","\$1Billion deal"], 
["OpenAI","Signed","in Texas"], 
["OpenAI","Signed","for AI partnership"], 
["OpenAI","Signed","in 2024"], 
["Microsoft","Partners\_With","OpenAI"], 
["Microsoft","Signed","\$1Billion deal"], 
["Microsoft","Signed","in Texas"], 
["Microsoft","Signed","for AI partnership"], 
["Microsoft","Signed","in 2024"] \\
\emph{Reasoning:} The set of triplets covers all core facts from the source text. \\
\emph{Warning:} None \\

\textbf{Score 2 (Partial):} \\
\emph{Set of Triplets:} 
["OpenAI","Partners\_With","Microsoft"], 
["OpenAI","Signed","\$1Billion deal"], 
["OpenAI","Signed","in Texas"], 
["OpenAI","Signed","for AI partnership"], 
["Microsoft","Partners\_With","OpenAI"], 
["Microsoft","Signed","\$1Billion deal"] \\
\emph{Reasoning:} It misses the date and does not cover all for Microsoft. \\
\emph{Warning:} Possible positional bias and missing information. \\

\textbf{Score 1 (Poor):} \\
\emph{Set of Triplets:} ["OpenAI","Partners\_With","Microsoft"] \\
\emph{Reasoning:} It misses core facts. \\
\emph{Warning:} Incomplete set of triplets generation. \\
\end{tcolorbox}